\documentclass[conference]{IEEEtran}
\usepackage{graphicx,subcaption}
\usepackage{pgfplotstable}
\usepackage{algorithm}
\usepackage{algpseudocode}
\usepackage{amsmath}
\usepackage{amsfonts,amssymb}
\usepackage{url}
\usepackage{subcaption}

\usepackage{amsthm}
\theoremstyle{definition}

\usepackage{multirow}
\usepackage{siunitx}
\usepackage{tikz} % To generate the plot from csv
\usepackage{pgfplots}

\pgfplotsset{compat=newest} % Allows to place the legend below plot
\usepgfplotslibrary{units} % Allows to enter the units nicely

\twocolumn

\usepackage{listings}
\usepackage{color}
\definecolor{lightgray}{rgb}{.9,.9,.9}
\definecolor{darkgray}{rgb}{.4,.4,.4}
\definecolor{purple}{rgb}{0.65, 0.12, 0.82}

\lstdefinelanguage{JavaScript}{
  keywords={typeof, new, true, false, catch, function, return, null, catch, switch, var, if, in, while, do, else, case, break},
  keywordstyle=\color{blue}\bfseries,
  ndkeywords={class, export, boolean, throw, implements, import, this},
  ndkeywordstyle=\color{darkgray}\bfseries,
  identifierstyle=\color{black},
  sensitive=false,
  comment=[l]{//},
  morecomment=[s]{/*}{*/},
  commentstyle=\color{purple}\ttfamily,
  stringstyle=\color{red}\ttfamily,
  morestring=[b]',
  morestring=[b]"
}

\ifCLASSINFOpdf
\else
\fi

\begin{document}

%\title{Proofware: A Crowd-based Elastic Autonomous Exascale Computing System}
\title{Proofware: Proof of Useful Work Blockchain Consensus Protocol for Decentralized Applications}

\author{
\IEEEauthorblockN{Zhongli Dong}
\IEEEauthorblockA{
School of Computer Science\\
The University of Sydney\\
Sydney NSW 2006, Australia\\
Email: zhongli.dong@sydney.edu.au}
\and
\IEEEauthorblockN{Young Choon Lee}
\IEEEauthorblockA{
Department of Computing\\
Macquarie University\\
Sydney NSW 2109, Australia\\
Email: young.lee@mq.edu.au}
\and
\IEEEauthorblockN{Albert Y. Zomaya}
\IEEEauthorblockA{
School of Computer Science\\
The University of Sydney\\
Sydney NSW 2006, Australia\\
Email: albert.zomaya@sydney.edu.au}
}

\maketitle

\begin{abstract}
In a blockchain system, consensus protocol as an incentive and security mechanism, is to ensure the participants to build the block honestly and effectively. There are different consensus protocols for blockchain, like Proof of work (PoW), Proof of Stake (PoS), Proof of Space (PoSpace), Proof of Activities etc. But most of these consensus protocols are not designed for doing some useful jobs for society because of too much competition and scalability limitation. Massive electric power and computing resources, including CPU, RAM, storage and sensors have been wasted to run blockchain network based on these consensus protocols. Current frameworks and middleware for building decentralised applications (dApps) are largely limited to simple and less useful jobs.
%easily based on public computing resources. That's the reason why most dApps is only for trading, gaming and gambling which is not cost much computing resourcing. 
Current blockchain frameworks, such as Ethereum, are difficult to support diversity of decentralised applications due primarily to redundantly performing `hashing'. 

In this paper, we present Proofware which is designed for developers to build their dApps easily with existing public/crowd-based computing resources. Under Proofware, developers can develop and test their own Proof of Useful Work (PoUW) consensus protocols. Also, rather than depending on a centralised accounting system, each dApp has an embedded currency system to keep the whole incentive system decentralised, fair, transparent, stable and sustainable. Based on Proofware, we have built a crowd based video sharing application, called OurTube, as a case study. By the OurTube example, it has shown Proofware significantly improves the productivity to build crowd-based computing system with the features of cost-effectiveness, anti-censorship, elasticity and financial sustainability.

Comparing with running the similar video application on Amazon EC2, the cost running on Proofware is only about 5.4\% of the cost running on Amazon EC2. As there is no centralised server or payment system, no one can stop the servers of OurTube or remove the contents from OurTube. In the experiment, by increasing the offer price, OurTube can attract unlimited participants to contribute their computing resources and also encourage content providers to share their video contents. As all the computing resource providers and content providers get proper payment, it keeps the system financial sustainability. More importantly, as we are using a decentralised price model, it shows the system can prevent market manipulation and control the stability of computing resource price and the supply volume of computing resource which can support the elasticity and scalability of OurTube application.  The experiment also showed the low and controllable transaction cost in our decentralised credit exchange model.

\end{abstract}

\


\IEEEpeerreviewmaketitle

\section{Introduction}
With proper consensus protocol, P2P network and decentralised applications(dApps), we could connect all worldwide computing resources together to build a crowd based large scale computing system. The computing resource here does not only refer to CPU, storage or network, also includes other input or output devices, like sensors, camera, GPS, etc. from massive personal mobile devices as well. A developer may want to build a reliable large distributed storage system in very short time with very limited cost. In the traditional way, they may need to buy storage server from server vendors and build their storage system. More conveniently, they could get these storage and computing resources from cloud service providers, such as Amazon Web Services (AWS), Microsoft Azure and Google Cloud Platform. In none of these ways, user’s requirements can be met as they either cost too high or need long time to build. A new way to build such system is to leverage spare storage spaces and computing resources which are distributed on worldwide personal systems, like home PC, or commercial system, like unused enterprise storage. Another scenario is that a mobile developer wants to build a mobile application to scan the bar code of product in the shop and upload the price to run a price comparison with other prices provided by other users. 

To build such system, the developer needs to design a good incentive mechanism and credit system to implement the application. Without a proper framework, it takes months to build such an application. In this paper, we present Proofware as a novel framework for dApp development. It is designed on the basis of Proof of Useful Work (PoUW). Proofware consists of: (1) A decentralised independent application currency and autonomous price adjustment design which can provide more effective, transparent and stable financial incentive system cross all applications, (2) A decentralised service contract to define the requirement and quality of the service (QoS), and (3) A flexible architecture can support different characteristic decentralised applications. 
%3. Crossing blockchain protocol to support interacting among different blockchain systems.
%4. An easy-to-use toolkit to build the DApp.

The first issue we need to resolve is to find an effective incentive and pricing mechanise. We need to find an effective way to ensure the public participant to contribute their devices to the network continuously. Bitcoin~\cite{nakamoto_bitcoin:_2008} is the most famous and successful blockchain project. Even some people think Bitcoin is dangerous speculative bubble, Bitcoin’s financial incentive really works well. In Bitcoin network, all participants, aka ``miners" have incentive to composite the block honestly with most power efficiency mining device. However, the Bitcoin's incentive system cannot cross the multiple applications. We need a decentralised incentive system which can support multiple applications independently and exchangeable.  The incentive system in our Proofware, learned from Bitcoin, adding the autonomous credit swapping feature and price discovery feature, use blockchain based credit system to reward all devices owners and incentive them to contribute their devices honestly to the network.

For the service contract, traditional cloud service providers define a list of services, list the price and define term and condition about their quality of service. But all these contracts are static, inflexible and difficult to audit. The service providers totally own these contracts.  The traditional contract life cycle needs human involvement which is easy to make a mistake and get disputed. In the Proofware system, the service providers are crowd-based, decentralised and dynamically, the traditional service contract cannot support such multi service providers and dynamic model effectively. We introduce decentralised service smart contract~\cite{buterin2014next}, running on blockchain, is a list of business rules to define the crowed decentralised service requirement details. The smart contract is decentralised, permanent and transparent which meet the decentralised service contract requirement. In our system, a decentralised service smart contract template will be provided so the developer can define these services easily. 

Proofware needs to support various applications which may need to invoke different devices. We use a `device adaptor' model to make our architecture flexible to support more types of computing devices and sensors without rewriting whole business logic code. Also, the component-based software engineering (CBSE) design principle makes each component could be replaced without involving much changing on other connecting components. Specific contributions of this paper are as follows:

\begin{itemize}
    \item We design a framework to support developing and running  dApps on a crowd-based large scale computing system
    \item We design an autonomous  credit swapping and decentralised pricing  mechanism to support more effective financial incentive for resource providers and more cost-controllable for dApp developers
    \item We verify the effectiveness of our credit exchange and pricing adjusting in a simulated environment
    \item We build OurTube as a demo dApp to show the cost-effectiveness and reliability of our Proofware system
\end{itemize}

%Such system also need to support cross-chain (?) trading functions to enable cross chain the token transactions. There are different tokens in the markets. The participants may need to use a token to buy other services which only accept their own issued token. Also, some service provider issue two type of token, one type is token an equity in the cryptocurrency exchange; another type of token is used in their application. These two types of token will need be exchanged when the user want to sell application to get equity token, and vice versa.

%A DApp toolkit which is included in our system is to help developers to develop these applications. This toolkit will include DApp development tools to help develop compilation, linking, deployment and binary management their DApp. It also helps developers to write these automated smart contract test case rapidly. The toolkit can include other smart contract or external packages. The toolkit also provides testnet docker image so developers can run testnet network and debug these DApp easily. 

The rest of this paper is organised as follows. In Section ~\ref{section:related_work}, we give background and review related work on the existing consensus protocol, blockchain credit system and cross chain protocols.  In Section ~\ref{section:architecture}, we describe the architecture of Proofware. In Section ~\ref{section:algorithms}, we present the algorithms of autonomous credit swapping and application credit price discovery which support our financial incentive mechanism, the core of Proofware.  We present experimental results in Section ~\ref{section:experimental} and our conclusion in Section ~\ref{section:conclusion}.

\section{Background and Related Work}
\label{section:related_work}
To reach consensus among each peers, typical public Blockchain consensus algorithms include Proof-of-Work (PoW)~\cite{laurie_proof--work_2004} used in Bitcoin and Ethereum, Proof-of-Stake (PoS)~\cite{king2012ppcoin} used in Peercoin and Delegated Proof-of-Stake (DPoS)~\cite{larimer2014delegated} used in EOS. In this section, we begin by describing key indicators that we think are essential for the success of public computing platform in comparison with existing other systems(Table ~\ref{table:compare}). We then discuss related technologies in relation with Proofware. 

\subsection{Key indicators for the platform to build public large scale computing platform}
\begin{itemize}
  \item Censorship Resistance and Self-regulation: The system should not be able to stopped by a centralised agency; The content of transported or stored cannot be removed or altered. Same time, the whole system should maintain all records transparently. Audit and validation could be done by public, not only by centralised authority.
  \item Decentralisation: The credit system, clients and the content should be decentralised. There is no centre controller  and no single point failure.
  \item Elasticity: The computing resource should be used as utility, on demand, rather than one-time purchase. 
  \item Autonomous: The system could be run with further reconfiguration or interruption.
  \item Throughput: The system can process the high volume of information with a liner scalability. 
  \item Reliability: The system has low failure rate. 
  \item Cost Effective: Both the setup cost and running cost should be lower comparing with centralised system.  
  \item Extensibility: The system would be able to extend to the different scenario of requirement.
 \end{itemize}
Based on the above system characteristics, we have compared our framework with the following existing computing paradigms:
\begin{itemize}
  \item Cloud Services: Cloud service is a centralised service to provide elastic computing resource in an utility mode with a low initial cost.
  \item Bitcoin Network: Bitcoin network provides decentralised open ledger service. Also, there are other blockchain networks which are not only play a ledger rule, but also run as decentralised general applications for other purpose. 
  \item Grid Computing: Grid computing~\cite{foster2003grid} uses a distributed task scheduler to split the tasks, distribute the tasks to the volunteer computers and merge the submitted result. Without an effective financial incentive and credit exchange, the participant rate of grid computing increases slowly.
  \item Crowdware~\cite{dong2015crowdware}: Similar with grid computing, but it added financial incentive mechanise to attract participants and auction mechanise to implement elastic price and resource matching.
  \item Proofware: the Proofware, we leverage blockchain based credit system an self-controlled incentive, autonomous smart contract for service and dApp toolkit to achieve all above key evaluation indicator.  
 \end{itemize}

\begin{table*}[!t]
\begin{minipage}{\textwidth}
\renewcommand{\arraystretch}{1.1}
\caption{Comparison of Large Scale Computing Platform.}
\label{table:compare}
\begin{tabular}{|c|c|c|c|c|c|c|c|c|}%{|p{1.5cm}|p{1.5cm}|p{1.5cm}|p{1.5cm}|p{1.5cm}|p{1.5cm}|p{1.5cm}|p{1.5cm}|p{1.5cm}|}
\hline

Name &  Censorship Resistance & Decentralisation   & Elasticity & Autonomous & Throughput & Reliability &Cost Effective & Extensibility \\ \hline
Karma  & Low            & Low                 & Medium       & Low       & Medium   & Medium   & High          & Medium        \\ \hline
Cloud Services  & Low            & Low                 & Medium       & Low       & Medium   & Medium   & High          & Medium        \\ \hline
Bitcoin Network & High           & Medium              & Medium       & Medium    & Low      & High     & Low           & Low           \\ \hline
%Polkadot       & Medium         & Medium              & High         & Low       & High     & Medium   & High          & High          \\ \hline 
%Dfinity       & High           & High                & High         & High      & High     & High     & High          & High          \\ \hline 
Grid Computing  & Low            & Medium              & High         & Low      & Low       & Medium   & High          & Low          \\ \hline 
Crowdware       & Medium         & Medium              & High         & Low       & High     & Medium   & High          & High          \\ \hline  Proofware       & High           & High                & High         & High      & High     & High     & High          & High          \\ \hline          
\end{tabular}
\end{minipage}
\end{table*}

\subsection{Consensus Algorithm}
\subsubsection{Proof of work(PoW)}
Proof of work (PoW) has been used extensively to prevent the DDoS attacks or bots attacks. When you do a web registration or a login, the web page always ask you to input a unrelated question or to do a job to verify you are a human, not a bot.  In the blockchain area, we use a similar way to verify every node which participates the block composition job is honesty each other, preventing the marvellous nodes. In the PoW algorithm, the piece of target data is difficult to produce but easy for others to verify, nodes in the network will compete against each other to solve a computational costed puzzle to generate the next block, as so called mining. The first node who solved the puzzle will broadcast its result in the network, then rewarded by the blockchain network if the result is verified by the majority of the peers. All of these guess work happen in peer locally despite of the scale of the network. Communication happens only when the puzzle is solved, therefore it is easy to scale out in a PoW-based blockchain network. Though PoW can solve the consensus and scalability problem, it costs a huge amount of computational resources (electricity) to maintain the security of the network and the energy consumption cannot be applied business, science or any other field~\cite{o2014bitcoin}. Besides, in the centralised mining factory and mining pool (i.e., hashrate distribution amongst mining pools \url{https://blockchain.info/pools}) of such distribution, the owner of mining pools can start 51\% attack and can control the feature of future feature of Bitcoin which may only benefit themselves.

\subsubsection{Proof of Stake (PoS)}
Proof of Stake (PoS), a concept initially proposed by Peercoin, unlike PoW, there is no mining process required for reaching consensus. PoS asks users to prove holding the certain amount of of currency. Thus, PoW eliminates the computing power usage of proofs of work.  PoS is designed to solve the disadvantage of PoW, in particular, high energy consumption. Instead of mining with costly energy computation, each node deposit amount of cryptocurrencies as their stake, the network then randomly pick a node to produce the next block regards the weight of each peer’s stake. However, PoS experience a huge drawback in scalability. As known, there is no centralised node for coordinating the stake among all the nodes, therefore, every node needs to participate in the election process, communication messages will grow up exponentially when the network scales out, which will consequently slow down the performance of the network.

\subsubsection{Delegated Proof of Stake (DPoS)}
To address the scalability problem in PoS network with unlimited nodes, some blockchain systems like Steem~\cite{larimer2016steem}, BitShares~\cite{schuh2015bitshares} and TRON~\cite{bashir2017mastering}, have adapted DPoS for their consensus protocol.  Comparing with PoS, the little volume token holders can vote their preferred node to express these opinion. In DPoS network, only a few nodes (usually an odd number) are participating in the election process, as the result, not every node consider themselves as a potential block producer, instead, they will delegate their stake to other nodes which have higher potential to be elected. Being benefited from a limited number of candidates, DPoS can reach a notable transaction per second (TPS). But there is bribing issue for voting. The candidates promise to share some portion of that as a bribe, equally split among all of their voters. There is also bad incentive to forming Cartels using the existing voting award mechanism. 

\subsubsection{Proof of Capacity}
Proof of capacity, also called proof of space~\cite{dziembowski2015proofs} where miners must dedicate a significant amount of disk space as opposed to computation in PoW. A typical implementation like Storj's Metadisk~\cite{wilkinson2014metadisk}, which is a decentralised cloud storage platform. It combines existing BitTorrent Sync~\cite{farina2014bittorrent}, Bitcoin and cryptography technologies.

\subsection{Blockchain Interoperability and Multi-chains Technologies}
One chain has limited performance and is not suitable for various business requirement. There are few blockchain interoperability solutions which try to connect multi blockchains to achieve high performance, different scenario and autonomous inter-chain communication.

\subsubsection{Coco Framework}
Coco framework (\url{https://github.com/Azure/coco-framework/}), from Microsoft, is trying to build an open and compatible framework with any blockchain protocol. Coco framework introduced a trusted execution environment (TEE) which is a trusted box on which users can trust to put their blockchain code. For the perspective of pluggable consensus protocol design, it is most similar with the framework we have build.
But it is literature only as we traced their github and didn't see any update after 16 August 2017.

\subsubsection{Cosmos}
Cosmos (\url{https://cosmos.network/}) is a network connecting many independent blockchains, called zones. The zones are powered by a main chain which is implemented by a Practical Byzantine Fault Tolerance (PBFT) consensus engine Tendermint Core~\cite{buchman2016tendermint}. Blockchains with other consensus models, including PoW blockchains like Ethereum and Bitcoin can be connected to the Cosmos network using adapter zones. The hub and zones of the Cosmos network communicate with each other via an inter-blockchain communication protocol. Tokens can be transferred from one zone to another through the Cosmos Hub, which keeps track of the total amount of tokens held by each zone. Cosmos also introduced ``fork-accountability", where the processes or malicious actors that caused the consensus to fail can be identified and punished according to the rules of the protocol. 

\subsubsection{Polkadot}
Polkadot~\cite{wood2016polkadot} connects permissioned blockchain, permissionless blockchain. It has introduced the concept of parallelisable chains, or called ``Parachains". Parachains are simpler forms of permissioned blockchains which gather and process transactions but without their own security protocols. These Parachains attach to a ``relay chain" which coordinates consensus and transaction delivery between chains. It also has ``Bridges" link to blockchains with their own consensus such as Ethereum.
how to prevent ``Tragedy of the commons''.

\subsubsection{Dfinity}
Dfinity~\cite{hanke2018dfinity} is working on decentralised cloud computing. Its ambition is to build a blockchain supercomputer, called cloud 3.0, designed to host the new decentralised of software and services. But Dinity only supports one quorum selections based consensus algorithm. Such single consensus is difficult to support different cloud computing requirement, like IoT cloud.

All these blockchain interoperable technologies just focus on to support existing consensus protocols. They do not provide a way for developer to define their own consensus protocols, the protocol for some proof of useful contribution. It also cannot provide the cross application credit system to support transfer the award between the different applications.

\section{Architecture}
\label{section:architecture}
In this section, we begin with the description of overall Proofware design. We then give the general blockchain framework stack. Finally, we introduce main roles and participants of Proofware platform.

\subsection{Overall}
\begin{figure}[!t]
\centering
\includegraphics[width=1\linewidth]{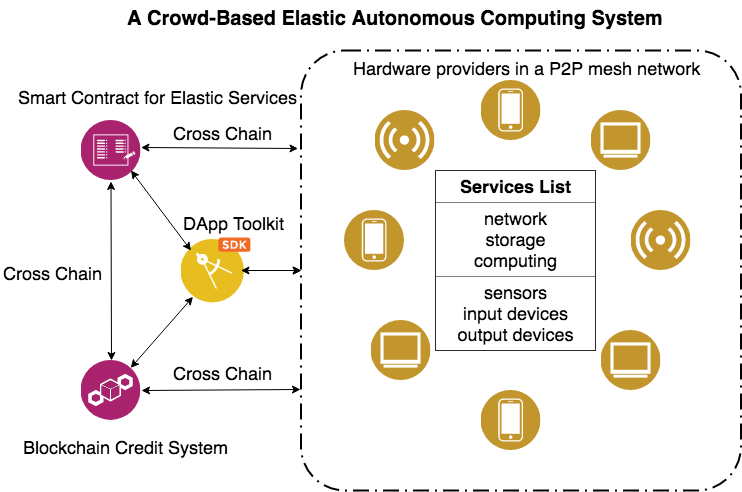}
\caption{The overview of Proofware.}
\label{figure:overview}
\vspace{-10pt}
\end{figure}

As shown in Figure ~\ref{figure:overview}, the Proofware platform provides cloud-like computing resources, storage, IoT devices, human input to help third-party DApp developers to operate their application on truly decentralised environment. At the same time, the Proofware platform also encourages third parties developers to develop a public service to connect all existing peer-to-peer source to a more friendly API so the dApp developer can leverage these public resources easily without losing the decentralised characteristic. dApps developers not only can use the public service infrastructure that Proofware platform already provides, but also can leverage these third-party public API services to accelerate their development process and reduce the operational cost.

\subsection{Design Principle}
\begin{itemize}
  \item Decentralised: In the design, we not only decentralise the computing node to implement a robust computing system, we also decentralise the financial system to achieve a stable incentive system for all participants. 
  \item Algorithmic Transparency: To build a crowed based trusted computing system, we put all algorithm logic in smart contract which  persistent the status to the blockchain permanently. 
  \item Elastic: The applications running on the Proofware should can consistently acquire enough computing nodes or power via autonomous price system.
  \item Low coupling and plug-able: The applications interact with the Proofware using gRPC protocol and each Proofware component can be replaced without involving changing on other components.
\end{itemize}

\subsection{General Blockchain Framework Stack}
\begin{itemize}
  \item Consensus protocols and algorithms: The consensus protocol or algorithm defines the way how the distributed nodes to achieve permanent and unanimous consistent.
  \item Software tokens and global addresses: Blockchain is a public open ledger, like a public account book.  To transfer the value on the blockchain, we need use token and address to represent the ownership of the value.
  \item Peer-to-peer based communication and storage: To keep decentralisation and robustness, the blockchain system should keep on the peer-to-peer network without centralised controller and without a single point of failure.
  \item API and Programming language support:
Blockchain system also needs to provide programming language support so the user can put their business logic on the script and these scripts can be interpreted and run on the blockchain system using the computing power of peer-to-peer nodes. The API is running above the blockchain layer and providing functions to interact with the blockchain network. It is designed to response requests from the official clients, performs as a bridge between the clients and the blockchain. It receives requests from the client side and response back after updating or querying the Blockchain. To achieve totally decentralisation, it could be run by any party in the ecosystem, either publicly or privately.
  \item Development tools Kit: Similar to the API, the developer tools are designed to interact with the consensus nodes with various methods. There are no official clients built on top of the developer tools. Developers can use the tools to configure and test their consensus nodes and develop their own dApps via public resources and their own consensus protocol.
  \item Transparent decentralised ledger and application credit system: We use public blockchain as transparent decentralised ledger. It is a distributed infrastructure that keeps immutable ledgers or footsteps in the network and enables other applications and tools to update and query against it. An unlimited number of nodes that run in any vendors are connected to each other via the Internet with a consensus algorithm running on it, which keeps the records consistent and immutable. 
\end{itemize}

\subsection{Main Roles and Participants of Proofware Platform}

\subsubsection{Public Service Developers}
Public service developers, using the provision of development frameworks and SDKs provided by the Proofware platform to provide public information services, using public service nodes or their own private resources to complete the development of public functions.
\subsubsection{Public Service Node Provider (miner)}
Public service nodes can be from cloud servers, PCs, mobile devices, IoT devices, and so on. The nodes could be classified into:
\begin{itemize}
  \item Consensus nodes: Responsible for the consensus of the public ledger, support the embedding and customisation of the consensus mechanism, and provide popular consensus mechanisms such as PoW, PoS, and DPoS by default.
  \item Information collection nodes: responsible for information collection, such as event result collection, public opinion and voting information collection.
  \item Computing nodes: Provides a decentralised high performance computing: e.g real time order matching system.
  \item Storage nodes and artificial intelligence (AI) analysis nodes: The storage node is responsible for a large amount of business data and other large file storage. Based on a large amount of data, these nodes also provide data analysis and AI functions.
  \item IoT nodes: Provides IoT information collection service. For example, integrated crowd-based smart city sensors to predict traffic conditions, forecast and manage major events.
  \item Arbitration nodes: used to verify the relevant results and adjust when disputes occur.
  \item Customised nodes. Based on business needs and scenarios, developers can define their own business nodes based on the template of the Proofware framework.
\end{itemize}

\subsubsection{dApp Developer}
dApp developers use the public service interface. They can take advantage of Proofware's existing incentives for dApp promotion and sustainable operation.
dApp developers can build their own computing networks, sub-chains, and consensus networks based on their business characteristics, and build their own incentive system based on Proofware's credit system.

\subsection{Proofware Platform Infrastructure}

\subsubsection{Template system}
\begin{itemize}
	\item Business process template: Provide several typical business process templates for dApp developers to reuse.
    \item Credit system template: It is convenient for dApp developers to create their own credit unit and implement the core of the decentralised business.
    \item Service Smart Contract Template: dApp developers can use the Proofware platform and third-party smart contract templates to define computing resource requirements, prices, quality of service and payment agreements.
\end{itemize}

\subsubsection{SDK}
\begin{itemize}
	\item Proofware main network SDK: detect the status of the main network, query transactions, initiate transactions.
    \item Third-party public services for interactive SDK: mainly used for convenient information exchange with third-party public service networks. dApp developers call other public resource implementations to implement dApp functionality
    \item Cross-chain SDK. Used by dApp developers to complete cross-chain transactions and cross-chain resource calls.
\end{itemize}

\subsubsection{Blockchain-based Credit Incentive System}
\begin{itemize}
	\item Accept and send Proofware credit and credits on other dApp subnets. It also provides the credit transfer and trading function crossing the applications.
\end{itemize}

\section{Algorithms}
\label{section:algorithms}
In this section, we present our main contribution on the two algorithms to construct the decentralised credit system for our Proofware platform. One algorithm is for autonomous credit swapping system and another one is for decentralised credit price discovery.

\subsection{Autonomous Credit Swapping Algorithm}
The Proofware credit, as incentives in decentralised applications, can be used to attract public service providers, access to services and as proof of contribution some useful jobs to the system. Proofware credit has been issued by the dApp developer via a smart contract.

An atomic swap (\url{https://en.bitcoin.it/wiki/Atomic_cross-chain_trading})，could be powered by a serial of  smart contracts, was used in trading between two peers who exchange different digital assets each other directly without depending on a third party or an escrow agency, more importantly, there is no default risk on two side peers. 

The main target for this algorithms as below
\begin{itemize}
    \item Instance: The credit should be swapping instantly to the user's own wallet without waiting for withdrawal from a centralised trading platform. 
    \item Minimum Transaction Fee: As there is no middle man during the credit swapping, it is no fee to pay for the transaction. But there is some transaction fee from the underlying blockchain. 
\end{itemize}

\begin{figure}[!t]
\centering
\includegraphics[width=1\linewidth]{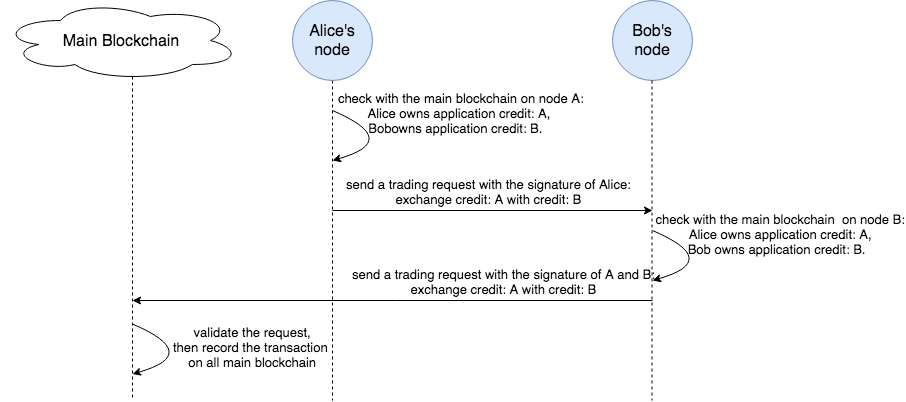}
\caption{Autonomous Decentralised Credit Trading.}
\label{figure:trading}
\vspace{-10pt}
\end{figure}

As showed transaction sequence in Figure ~\ref{figure:trading}, Alice and Bob completed there application credit swapping without any third-party agency. All the transaction process has been done on the blockchain with their own signed transaction details.

\subsection{Decentralised Price Discovery Algorithm}

Decentralised price discovery ~\cite{ranjan2007decentralised} refers to the act of determining the proper price of service by calculating market supply and demand without centralised broker service. Discontinued the market liquidity from the trade value. The Users can always exchange their earned credit units to the different credit units via smart contracts even when there are only few or no other buyers or sellers in the market.

We formulate the problem as the following:
\begin{description}
  \item[Main Credit Unit] \hfill \\ Main credit unit is a token which is directly exchangeable with other token.
  \item[Application Credit Unit] \hfill \\ Application credit unit is a token issued by the application developer for the DApp. 
  \item[Credit Adaptor] \hfill \\ Credit adaptor holds a balance of another credit to which it is connected.
\end{description}
\begin{itemize}
  \item Let $W_a$ denote the weight of a credit adaptor
  \item Let $B_a$ denote the application credit adaptor's balance
  \item Let $T_m$ denote the total value of total main credit units
\end{itemize}

\begin{equation}
W_a = \dfrac {B_a}{T_m}
\end{equation}

\begin{itemize}
  \item Let $P_m$ denote the main credit price
  \item Let $N_m$ denote the number of total main credit units
\end{itemize}

\begin{equation}
T_m = P_m \times N_m
\end{equation}

\begin{itemize}
  \item Let $W_i$ denote the weight of adaptor $A_i$
  \item Let $O_m$ denote outstanding supply of main credit units.
  \item Let $P$ denote the price of application credit unit.
\end{itemize}

\begin{equation}
P = \dfrac {B_a}{O_m \times W_a}
\end{equation}

\begin{itemize}
  \item Let $CE_a$ denote coefficient of supply of application credit unit
  \item Let $T_a$ application tokens have been paid
  \item Let $S_a$ application tokens supplied
\end{itemize}
\begin{equation}
CE_a = (1 + \dfrac{T_a}{B_a})^{W_a} - 1
\end{equation}
\begin{equation}
T_a = S_a \times CE_a
\end{equation}

\begin{itemize}
  \item Let $Paid_a$ denote paid out of application credit unit
  \item Let $CE_p$ denote coefficient of application credit paid out
  \item Let $Paid_a$ denote application credit paid out
\end{itemize}
\begin{equation}
CE_p = \sqrt[W_a]{1+\dfrac{DE_m}{B_a}} - 1
\end{equation}
\begin{equation}
Paid_a = B_a \times CE_p
\end{equation}

Based on above formula, we leverage main credit as a media which provide continuous liquidity by incorporating an autonomous application credit market making functionality directly into their smart contracts. Main credit use application credit's balance and formula to offer to trade application credits at a predictable prices.

\section{Experiments and Results}
\label{section:experimental}
In this section, we first present our experimental setup and then we show our framework's price discovery API and and smart contract API; then we implemented a decentralised application to share the video between peers via public storage facility, called OurTube. Preliminary experiment has been OurTube on system with 40 virtual PCs connected. We also run the same video playing workload on a centralised 4 nodes Amazon EC2 cluster. Then we compare the cost and reliability between OurTube and centralised solution.

\subsection{Experimental Setup}

Table ~\ref{table:pc_config} shows our experimental hardware configuration and software configuration on our local hardware to simulate crowd-based hardware and on Amazon EC2 instances to show to performance and cost of centralised cloud environment.

\subsection{Price Discovery API Implementation}

\subsubsection{Price Discovery} 

\begin{itemize}
\item  URL: \url{https://localhost/1.0/price}

\item Request:
\begin{lstlisting}
{
"MainAppCredit":"proofware",
"fromAppCreditId":"91f33e6a-b7bc-11e8-96f8-529269fb1459",
"toAppCreditId":  "1f8d70bf-96d4-4996-ac19-2e154cd2531f",
"amount":"500",
}
\end{lstlisting}

\begin{itemize}
  \item MainAppCredit: Proofware credit
  \item fromAppCreditId: A unique application credit type ID that will indicate the application credit type the user would like to convert from. 
  \item toAppCreditId: A unique application credit type ID that will indicate the application credit type the user would like to convert to.     
  \item amount: the smallest application credit unit you would like to convert.
\end{itemize}

\item Response:
\begin{lstlisting}
{"price":"68"}
\end{lstlisting}
\begin{itemize}
  \item price: the amount of application credit unit (toAppCreditId) the user could receive for one application credit unit (fromAppCreditId).
\end{itemize}

\end{itemize}

\subsubsection{Transaction}

\begin{itemize}

\item URL: \url{https://localhost/1.0/swap}

\item Request:

\begin{lstlisting}
{
"blockchainType":"proofware",
"fromAppCreditId":"91f33e6a-b7bc-11e8-96f8-529269fb1459",
"toAppCreditId":  "1f8d70bf-96d4-4996-ac19-2e154cd2531f",
"amount":"500",
"ownerAddress":"0x0788a4c4Ff559eCe29C0a670E0eD76cB3e626511"
}
\end{lstlisting}

\item Response:
As we built our main blockchain component  based on Ethereum~\cite{wood2014ethereum}, the transaction cost is counted by gas which is the execution fee for every operation made on Ethereum network. Gas price refers to the amount of Ethereum the submitter of transaction to pay for every unit of gas. The transaction cost equals to gas times gas price.
\begin{lstlisting}
{
"data": [
{
"from": "0x0788a4c4Ff559eCe29C0a670E0eD76cB3e626511",
"to": "0x51250A16500C19c72828B21f15F42Ee52ca11d51",
"data": "0xf0843ba90000000000000000000000000000000da4c7bd
0f1de226d9e",
"value": "0x500",
"gasPrice": "0x28587600",
"nonce": "0x118",
"gasLimit": "0x8ab33"
}
]
}
\end{lstlisting}
\end{itemize}
After receive the response, the node can send the composited transaction to the main blockchain.

\subsection{Decentralised Application Smart Contract API}

Similar with Loom network (\url{https://loomx.io/}), We implemented smart contract with plugin model in GO language. Via implemented this interface, the dApp developer can easily issue their own application credit and elastic service smart contract. 
\begin{lstlisting}
import (
	proofware "./interface"
	token "./token_template"
	contract ./contract_template"
)
\end{lstlisting}

\begin{table}[!t]
\renewcommand{\arraystretch}{1.1}
\caption{Experiential Environment}
\label{table:pc_config}
\begin{tabular}{|p{2cm}|p{5.7cm}|}
\hline
Server       & 2 X HP Z420 Hexa-Core Workstations                                                          \\ \hline
Storage      & 10 X Seagate Backup Plus Desk Hub 8TB                                                    \\ \hline
RAM          & 128 GB DDR3 1333MHz                                                        \\ \hline
Virtualisation        & Xen 4.10.1 \\ \hline
Virtual Machines          & 40 X Ubuntu 18.04.1 LTS Desktops on local HP server                                                   \\ \hline
Amazon EC2           & 4 X Ubuntu 18.04.1 LTS Desktops, Instance Type: t3.large                                                   \\ \hline
Main Blockchain & based on go-ethereum Swarming (v1.8.13)                                       \\ \hline
Applications Blockchains &  Smart Contracts in Go Language                                      \\ \hline
Experimental Application &  OurTube, A decentralised video sharing system based on Proofware framework                                      \\ \hline
\end{tabular}
\vspace{-10pt}
\end{table}

\begin{figure}[!t]
\centering
\includegraphics[width=1\linewidth]{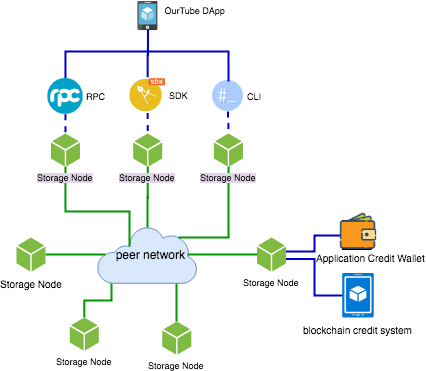}
\caption{Decentralised OurTube Application Running on the Proofware.}
\label{figure:OurTube}
\vspace{-10pt}
\end{figure}

\subsection{Experimental Application Choice}

One of the most exciting decentralised application is an decentralised on-line video sharing system with the financial incentive. Unlike YouTube and Youku, the content has been hosted by a centralised server by one organisation. In a decentralised video sharing system, the content is provided and controlled by the crowd. Same time, the censorship policy and rule can be defined by crowd via smart contracts. It's also easy to build-in autonomous financial incentive system and trusted review system.

OurTube is an experimental live video sharing network protocol that is decentralised, elastic, autonomous and cheaper than using traditional centralised or cloud based live video sharing solutions. The basic feature of OurTube is as below:

\begin{itemize}
	\item Broadcast a live stream into the peer-to-peer network and store the video to the crowd-based storage system. The developer or consumer pay the fee and content providers and storage service providers will get reward.
    \item The content provider or consumer can request that the video stream be encoded into multiple formats and size.
    \item The customers consume live stream from the any peer node and pay the expense via their application credit wallet.
\end{itemize}

\begin{itemize}
\item High Scalability
OurTube is based on a peer-to-peer network which allows develop to scale to any size without a single point of failure. The peer nodes also can  encode the video to the format which is best for the viewer's device.
\item Low Cost
Comparing with centralised cloud solution, like AWS, OurTube aims to reduce the price of storage, transport, encoding and transcoding by leveraging the resource of crowd node with the blockchain based finical incentive mechanisms.
\item Decentralised
No single organisation can control or run the censorship for the content of video. All the credit gained by content providers and reviewer can be traded freely global. Also, via our price decentralised price discovery algorithm, no market marker can manipulate the resource price.
\end{itemize}

\subsection{Results}

\subsubsection{Feature Comparison}
As table ~\ref{table:feature_compare}, we compare the feature of OurTube and CloudTube. It shows the advantages of from censorship resistance, elasticity and autonomous comparing with YouTube. 

\begin{table}[!t]
\renewcommand{\arraystretch}{1.1}
\caption{OurTube and YouTube Feature Comparison}
\label{table:feature_compare}
\begin{tabular}{|p{2cm}|p{2.5cm}|p{2.5cm}|}
\hline
Feature      & OurTube     &               CloudTube                                     \\ \hline
Censorship Resistance      & Yes     &               No                                     \\ \hline
Decentralisation      & Yes     &               No                                     \\ \hline
Elasticity      & Yes     &               Need Infrastructure support                                     \\ \hline
Autonomous      & Yes     &               No                                     \\ \hline
\end{tabular}
\vspace{-10pt}
\end{table}

\subsubsection{Benchmark Comparison}
Because of lack of YouTube platform's running data, it is difficult to benchmark the feature from quantitative perspective. 
Based the Amazon EC2 platform we build a centralised solution via NGINX web server and its RTMP module. We call this solution as CloudTube because it is running on AWS cloud.

In this experiment, we ran 1000 viewer clients to view the 2 weeks video stream same time to further verify the reliability of OurTube in terms of failure rate of executions. 
%Figure ~\ref{figure:benchmark} shows our crowd based OurTube uses about 5\% cost, comparing with Amazon EC2, implemented the similar reliability and latency.
%In the two weeks experimental, our AWS cloud's bill is about \$128.5. In particular, because OurTube is using the cheap public computing network resource and storage resource, only cost about \$7 Australia dollar achieved similar reliability.
We compare the performance of centralised CloudTube and crowed-resource based OurTube from the following 3 perspectives:
\begin{itemize}
  \item Reliability: OurTube achieved 99.5\% reliability, comparing with CloudTube's 99.7\% reliability which is running on Amazon EC2 instances. 
  \item Cost: The OurTube uses about 5\% cost of CloudTube which is running on Amazon EC2 instances. AWS cloud's bill is about \$128.5 Australian dollars. OurTube only costs about 7 Austrlian dollars under our simulated environment. 
  \item Latency: We found it is a noticeable delay during the initial phase of connection for OurTube. But it will become normal after the node found more and more peer nodes in the peer-to-peer network. Overrall, OurTube achieved 240.5 millisecond delay, comparing with CloudTube's 136 millisecond delay, it is pretty good reslult for a crowd based solution.   
\end{itemize}

\subsection{Simulated Application Credit Swapping Price Analysis}
\begin{figure}[!t]
  \begin{center}
    \begin{tikzpicture}[scale=0.7]

\pgfplotsset{
 scale only axis,
 %scaled x ticks=base 10:1,
 xmin=0, xmax=16,
}
\begin{axis}[
%title=Simulated Auction,
xtick={0,...,16},ytick={1, ..., 8},
axis y line*=left,
ymin=0, ymax=8, grid=both,
xlabel=Time,
x unit=\si{\hour},
y unit=\si{in Main Credit Unit},
ylabel=Trading price of OutTube Credit,
y tick label style={
        /pgf/number format/.cd,
        fixed,
        fixed zerofill,
        precision=0,
        /tikz/.cd
    },
]
\addplot[smooth,red,dashed]%mark=x,
table[col sep=comma,trim cells=true, x=time,y=w10]{data/simulated.csv};
\addlegendimage{/pgfplots/refstyle=plot_one}
\addlegendentry{Weight of Adaptor = 1}
\addplot[smooth,blue]%,mark=*
table[col sep=comma,trim cells=true, x=time,y=w5]{data/simulated.csv};
\addlegendentry{Weight of Adaptor = 0.5}
\addplot[smooth,dotted, black,mark=*,mark options={scale=0.5}]%,mark=*
table[col sep=comma,trim cells=true, x=time,y=w8]{data/simulated.csv};
\addlegendentry{Weight of Adaptor = 0.8}

\label{plot_one}
\end{axis}

    \end{tikzpicture}
    \caption{Application Credit Swapping Price During 16 Hours.}
    \label{figure:swapping_price}
  \end{center}
\vspace{-10pt}
\end{figure}
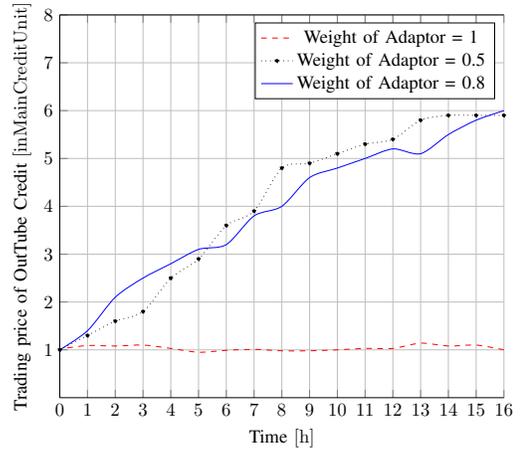

During our experiments, to verify our algorithms for autonomous credit swapping and the effectiveness of decentralised credit price discovery, we launch a 500 trading bot to swap OurTube credit during the 16 hours using Proofware's credit system in 3 rounds. Each round, we have set different weight of credit adaptor to 1, 0.5 and 0.8 respectively. We found the price elasticity and the price curves are alignment with the expectation of algorithms. During our testing, we have set any initial supply 500 and the start price as \$1 for each OurTube credit. The result shows in figure ~\ref{figure:swapping_price}:

\begin{itemize}
\item Set weight of credit adaptor to 1:
 OurTube's credit unit's price changes is very little with mail credit unit. The price is effectively pegged to its credit adaptor's balance. The mail credit plays a proxy for the value of OurTube's credit unit.
\item Set weight of credit adaptor to 0.5:
The OurTube's price moves linearly with the time and supply which is growing. The OurTube's appliction credit unit price decreases when demand for it is low and increases when demand for it is high. This relationship is matched with the law of demand states.
\item Set weight of credit adaptor to 0.8:
OurTube credit unit’s price reacts less and less to changes in supply in this case which is expected as the weight of credit adaptor is more than 50\%. 
\end{itemize}
Via above simulations, it shows our price discovery algorithm and is consistently determine their own reliable and predictable prices, which is essential for the mass adoption of application credit unit and attracting more resource contributors to our Proofware network.

\subsection{Decentralised Credit Exchange Transaction Analysis}

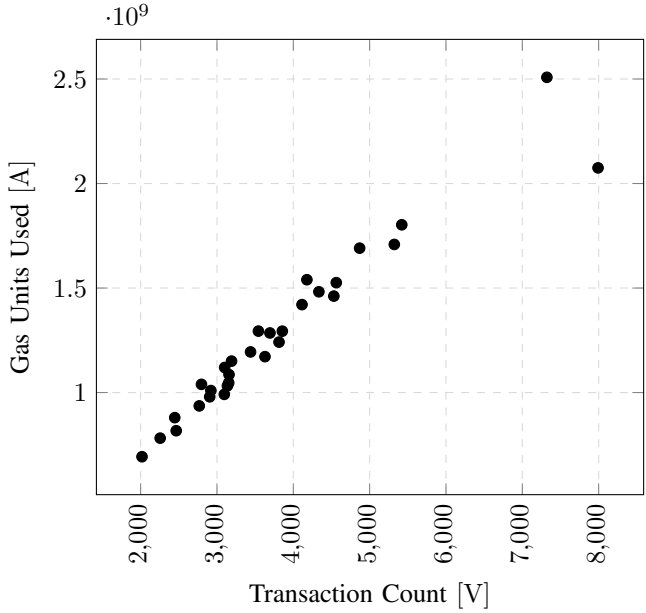
\begin{figure}[h!]
  \begin{center}
    \begin{tikzpicture}
      \begin{axis}[
          width=\linewidth, % Scale the plot to \linewidth
          grid=major, % Display a grid
          grid style={dashed,gray!30}, % Set the style
          xlabel=Transaction Count, % Set the labels
          ylabel=Gas Units Used,
          x unit=\si{\volt}, % Set the respective units
          y unit=\si{\ampere},
          legend style={at={(0.5,-0.2)},anchor=north}, % Put the legend below the plot
          x tick label style={rotate=90,anchor=east} % Display labels sideways
        ]
        \addplot[only marks]
        % add a plot from table; you select the columns by using the actual name in
        % the .csv file (on top)
        table[x=tx_count,y=sum_receipt_gas_used,col sep=comma] {data/Nov_Sum.csv}; 
        %\legend{Receipt Gas Used}
      \end{axis}
    \end{tikzpicture}
    \caption{Transaction Count and Gas Units Used}
    \label{figure:tran_count_gas_used}
  \end{center}
\end{figure}

As shown in the Fig. ~\ref{figure:tran_count_gas_used}, through 31 days simulation running, the daily transaction volume is from 2000 to 8000, the daily units of gas used is from \num{0.3e-9} to \num{2.5e-9} units which is matched the change trends of the number of daily transactions.

\begin{figure}[h!]
  \begin{center}
    \begin{tikzpicture}
      \begin{axis}[
          width=\linewidth, % Scale the plot to \linewidth
          grid=major, % Display a grid
          grid style={dashed,gray!30}, % Set the style
          xlabel=Transaction Count, % Set the labels
          ylabel=Transaction Fee in $Ether$,
          x unit=\si{\volt}, % Set the respective units
          y unit=\si{\ampere},
          legend style={at={(0.5,-0.2)},anchor=north}, % Put the legend below the plot
          x tick label style={rotate=90,anchor=east} % Display labels sideways
        ]
        \addplot[only marks] 
        % add a plot from table; you select the columns by using the actual name in
        % the .csv file (on top)
        table[x=tx_count,y=sum_receipt_gas_fee_paid,col sep=comma] {data/Nov_Sum.csv}; 
        %\legend{Paid Transaction Fee}
      \end{axis}
    \end{tikzpicture}
    \caption{Transaction Count and Transaction Fee}
    \label{figure:tran_count_trans_fee}
  \end{center}
\end{figure}
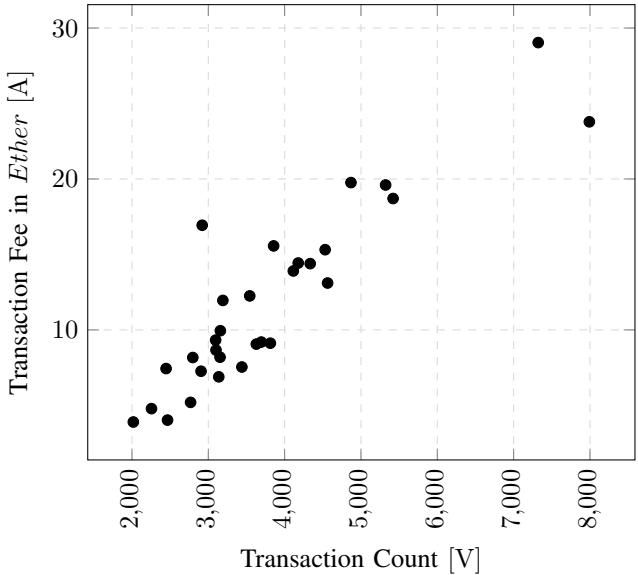

As shown in the Fig. ~\ref{figure:tran_count_trans_fee}, similar with the relationship between the the volume of transactions and gas unit used,the daily total transaction fee ranges from 3 to 29 ethereum which is also matched the change trends of the number of daily transactions. Note as the transaction fee is calculated based on gas units used and gas price, gas price is dynamic value which depends on computing power of Ethereum network, the transaction fee and gas units used have little different distribution.

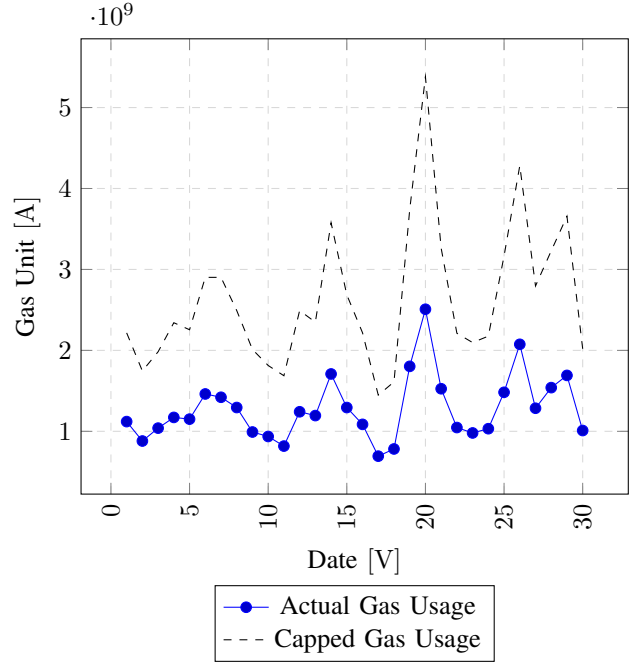
\begin{figure}[h!]
  \begin{center}
    \begin{tikzpicture}
      \begin{axis}[
          width=\linewidth, % Scale the plot to \linewidth
          grid=major, % Display a grid
          grid style={dashed,gray!30}, % Set the style
          xlabel=Date, % Set the labels
          ylabel=Gas Unit,
          x unit=\si{\volt}, % Set the respective units
          y unit=\si{\ampere},
          legend style={at={(0.5,-0.2)},anchor=north}, % Put the legend below the plot
          x tick label style={rotate=90,anchor=east} % Display labels sideways
        ]
        
        \addplot
        % add a plot from table; you select the columns by using the actual name in
        % the .csv file (on top)
        table[x=tx_date,y=sum_receipt_gas_used,col sep=comma] {data/Nov_Sum.csv}; 
        %\legend{Receipt Gas Used}
        \addlegendentry{Actual Gas Usage}

        \addplot[dashed]
        % add a plot from table; you select the columns by using the actual name in
        % the .csv file (on top)
        table[x=tx_date,y=sum_gas,col sep=comma] {data/Nov_Sum.csv}; 
        \addlegendentry{Capped Gas Usage}
        %\legend{Receipt Gas Used}        
      \end{axis}
    \end{tikzpicture}
    \caption{Actual Gas Usage and Capped Gas Usage}
    \label{figure:actual_capped_gas_usage}
  \end{center}
\end{figure}

The Fig. ~\ref{figure:actual_capped_gas_usage} shows the daily actual gas usage and capped gas usage. The capped gas usage to miner is more than the actual paid gas. It means the transaction cost is predictable which is necessary condition for a decentralised credit exchange system.

\begin{figure}[!t]
\centering
\includegraphics[width=1\linewidth]{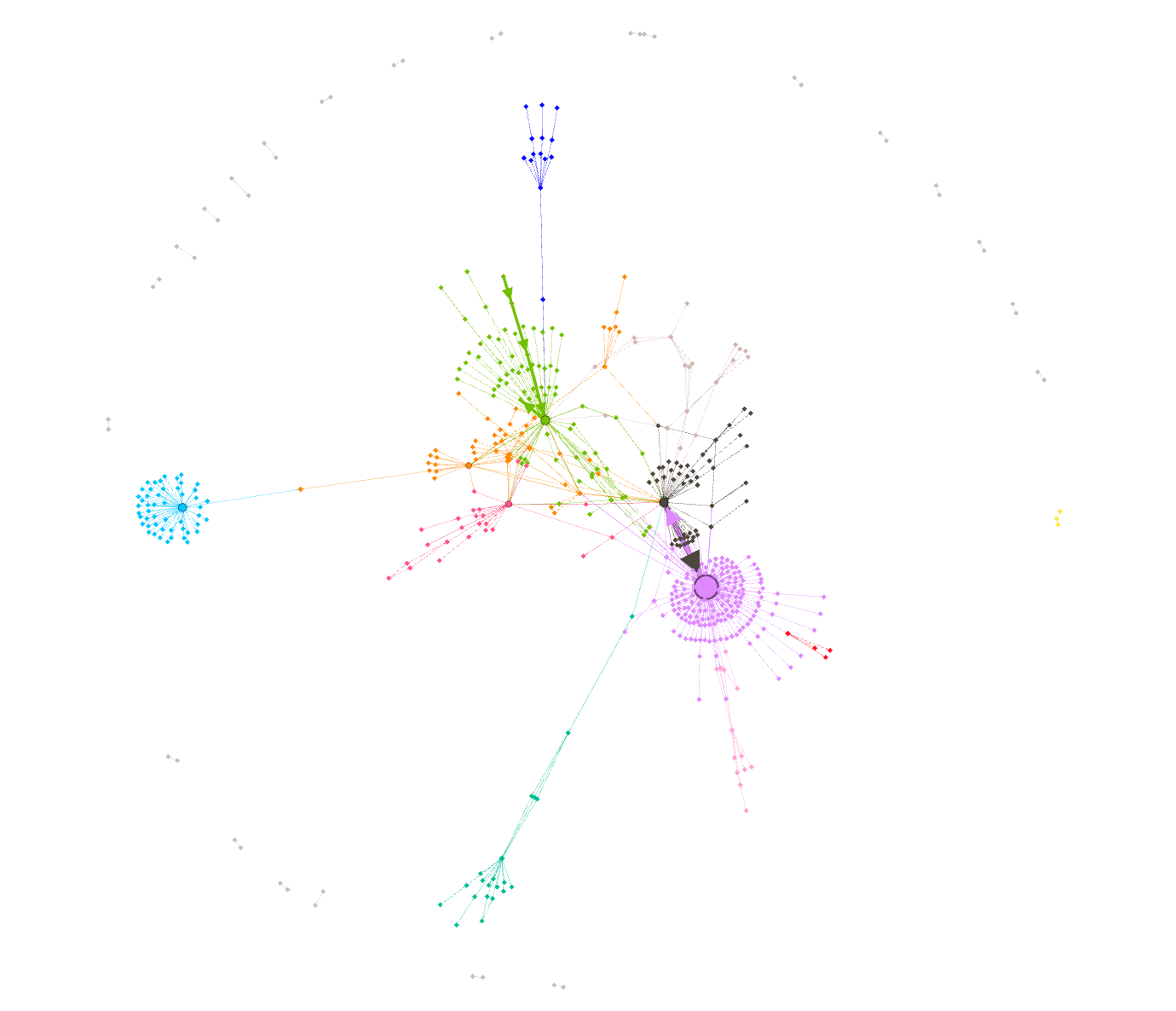}
\caption{Transaction Modularity}
\label{figure:transaction_modularity}
\vspace{-10pt}
\end{figure}

The Fig. ~\ref{figure:transaction_modularity} shows the modularity of the credit exchange transaction in the platform. From the modularity diagram, we could find the credit exchange network communities. The communities are groups by the different dApp credit. The unique address of credit holder make us the extract such information from a decentralised blockchain platform.

\section{Conclusion}
\label{section:conclusion}
Our Proofware provides the ability to efficiency harness crowd computing to build an elastic autonomous large scale computing System. In the Proofware, the dApp developer can build their own application easily via smart contract template and application credit incentive to attract public computing resource or human resources. Comparing with traditional computing resource, we have devised an effective price discovery mechanisms and automatic credit swapping algorithm to improve whole system's transparency and autonomacy to prevent corruption and price manipulation. Together, via blockchain and smart contract, Proofware significantly promotes crowd-based computing. Our results verify our claims with good profitability and cost efficiency in comparison with a cloud computing approach to build a large scale computing system with decentralised features.

Unlike the private computing system or virtual private cloud, blockchain is a public system. We plan to further investigate how to protect privacy in the crowd based public decentralised computing system. 

%\section*{Acknowledgement}
%Dr. Young Choon Lee would like to acknowledge the support of the Australian Research Council Discovery Early Career Researcher Award (DECRA) Grant DE140101628.

\ifCLASSOPTIONcaptionsoff
  \newpage
\fi

\bibliographystyle{IEEEtran}
%\bibliography{Zotero}  
\bibliography{proofware}

\end{document}